# The suppression of Finite Size Effect within a Few Lattices

Tao Liu[1], Kai Bai[1], Yicheng Zhang[1], Duanduan Wan[1†], Yun Lai[2], C.T. Chan[3], and Meng Xiao[1, 4*]

[1]Key Laboratory of Artificial Micro- and Nano-structures of Ministry of Education and School of Physics and Technology, Wuhan University, Wuhan, China

[2]National Laboratory of Solid State Microstructures, School of Physics, and Collaborative Innovation Center of Advanced Microstructures, Nanjing University, Nanjing, 210093, China

[3]Department of Physics, The Hong Kong University of Science and Technology, Clear Water Bay, Kowloon, Hong Kong, China

[4]Wuhan Institute of Quantum Technology, Wuhan 430206, China

Corresponding E-mail: † ddwan@whu.edu.cn; * phmxiao@whu.edu.cn

**Abstract**: Boundary modes localized on the boundaries of a finite-size lattice experience a finite size effect (FSE) that could result in unwanted couplings, crosstalks and formation of gaps even in topological boundary modes. It is commonly believed that the FSE decays exponentially with the size of the system and thus requires many lattices before eventually becoming negligibly small. Here we identify a special type of FSE of some boundary modes that apparently vanishes at some particular wave vectors along the boundary. Meanwhile, the number of wave vectors where the FSE vanishes equals the number of lattices across the strip. We analytically prove this type of FSE in a simple model and prove this peculiar feature. We also provide a physical system consisting of a plasmonic sphere array where this FSE is present. Our work points to the possibility of almost arbitrarily tunning of the FSE, which facilitates unprecedented manipulation of the coupling strength between modes or channels such as the integration of multiple waveguides and photonic non-abelian braiding.

# I. INTRODUCTION

A bound state can be trapped by a barrier. When the width and height of this barrier are not infinitely large, there is some probability that the state can tunnel through this barrier. These size-dependent phenomena are commonly called the finite size effect (FSE). FSE is ubiquitous for both quantum and classical waves. The barrier here can be a potential barrier such as a quantum dot[1,2] or originates from a band gap material[3,4]. Besides the intrinsic absorption loss, the probability of tunneling determines the lifetime of these trapped states, which is crucial for quantum information processing[5-8]. Control over the tunneling probability also enables manipulating interaction between different trapped states to generate various entangled quantum states[9-11]. Meanwhile, fine-tuning of the coupling coefficient is a key requirement in programmable photonic simulators[12], non-abelian braiding of photons[13] and quantum computers[14], especially when nanostructure are considered.

Similar to trapped states, propagating states such as the waveguide modes also exhibit FSE[15]. The tails of waveguide modes extend outside the waveguide with a length scale characterized by the penetration depth. If two waveguides are placed within the penetration depth of each other, there will be unavoidable intercoupling. The crosstalk between waveguides limits the integration of multiple waveguide channels into a compact device[16]. Boundary (hinge) modes localized on the boundaries of a strip (hinge of a bar) geometry of a periodic lattice can also be regarded as waveguide modes. With the recent explosive growth of research in topological physics[17-20], topological boundary modes and hinge modes associated with nontrivial bulk topologies have attracted a lot of attention due to their robustness against disorder and fabrication imperfections[21-25]. However, these boundary and hinge modes also suffer from the FSE when the width of the system is not large enough[26-29]. So, even topological edge modes can be gapped if the width is not big enough to stop the coupling of modes localized on the opposite edges of the sample. On the other hand, controlling the FSE of topological boundary modes or hinge modes can achieve new versatile controllability, such as spin flipping[30], electrical switching[31], *etc*.

In this paper, we present a novel type of FSE of boundary modes. Without loss of generality, we consider the boundary modes on a two-dimensional (2D) strip geometry. The boundary modes localized on opposite sides (if they exist) interact with each other, giving rise to an FSE-induced gap. Different from the prevailing understanding that the FSE vanishes only when the width of the strip is large enough, we discovered that the FSE could vanish at specific wave vectors (nodes) in a narrow finite-width strip. Interestingly, the number of nodes equals the number of lattices perpendicular to the boundary, i.e., the width of the strip. Using a model Hamiltonian, we analytically solved this system and proved the existence of this novel feature, and then demonstrated a filtering effect by utilizing this unique feature. Our system consists of only in-plane dipole orbitals and hence should represent general physics. Moreover, we provide a physical system composed of a plasmonic sphere array where this peculiar FSE can be observed.

The remainder of this paper is organized as follows. In Sec. II, we compare the difference between the typical FSE and this special type of FSE. We provide a tight-binding model Hamiltonian based on coupled in-plane dipoles $P_x$ and $P_y$, and further show the feature of the number of nodes equals the number of lattices along the y-direction ($N_y$). In Sec. III, we present the filtering effect by using this unique feature to filter out the component we need. In Sec. IV, we show that this peculiar type of FSE also exists in a plasmonic sphere array where all the coupling terms are taken into consideration. Hence our model is not limited to the tight-binding Hamiltonian. Finally, we summarize in Sec. V.

## II. MODELS OF THE PECULIAR TYPE OF FSE

Our system is shown in Fig. 1(a) which is periodic along the *x*-direction and finite along the *y*-direction. Each unit cell contains one atom or meta-atom which can support multiple modes such as $S$, $P_x$, $P_y$, $P_z$, *etc*. These modes interact and evolve into bands in momentum space and the physics can be captured succinctly using the usual tight-binding description. Inside a band gap, the system may exhibit boundary modes if the parameters are appropriately chosen, as shown schematically in Fig. 1(a) with the red and blue shaded regions denoting the mode profiles. These boundary modes can either be Shockley states[32], Tamm states[33] or originate from topological reasons[17,18,20,21]. In Fig. 1(b), we sketch the typical consequences of the FSE for a strip geometry of the system. Here the shaded areas represent the projection of bulk bands, and the blue and red lines represent the dispersion of the boundary modes. If there is no symmetry that forbids the boundary modes to couple, they will couple when they cross each other, always forming mini-gaps. However, if the system possesses mirror symmetry ($m_y$) as in our case, the boundary modes on opposite sides of the strip should exhibit the same dispersion when the strip is wide enough. Thus, for a finite size, there will be an interaction between boundary modes and once again form a mini-gap. In other words, the presence of a min-gap as shown in Fig. 1(b) seems unavoidable.

In contrast to Fig. 1(b), we present here a special type of FSE as pictorially shown in Fig. 1(c). Due to the mirror symmetry, the boundary modes can couple to form even and odd modes, as shown respectively by the red and blue lines. They twist with each other and form several nodal points. As will be shown later, the FSE of the boundary modes vanishes at these nodal points (no interaction between boundary modes). Intriguingly, we find that the number of nodes equals the number of lattices along the $y$-direction ($N_y$). Anomalous FSEs has been noted for helical boundary modes in topological insulator where the strength of the FSE decrease non-monotonically with the size[28,29,34], therein the oscillation length is hundreds or thousands of lattice constant. Therefore, the anomalous FSE is significantly different from our case as the FSE here oscillates at the lattice scale. Meanwhile, the two boundary modes in our case are related by mirror symmetry and thus exhibit the same dispersion in the absence of FSE. In contrast, the helical boundary modes in Refs. [26,28,29] exhibit opposite group velocities.

First, we provide a tight-binding model Hamiltonian that exhibits the salient features outlined in Fig. 1(c). We derive this Hamiltonian with the coupled dipole equation[35]. We assume each site supports two dipoles $P_x$ and $P_y$. For simplicity, we assume that other excitations are either far away in energy or are orthogonal to $P_x$ and $P_y$ (e.g., the $P_z$ dipole). The hopping from a dipole $\mathbf{P}_j$ at lattice site $\mathbf{R}_j$ to the dipole $\mathbf{P}_i$ at $\mathbf{R}_i$ is given by[36]

$$\ddot{\mathbf{G}}(\mathbf{r}) = -\frac{t}{r^5}\left[3(\mathbf{P}_j \cdot \mathbf{r})\mathbf{r} - r^2\mathbf{P}_j\right], \qquad (1)$$

with $\mathbf{r} = \mathbf{R}_j - \mathbf{R}_i$, and here $t$ is introduced only to ensure that the unit of hopping is energy. The "−" sign is intentionally introduced here to match the latter simulation of a real system. For convenience, we set $t=1$ and $r$ is in a unit of the lattice constant $a$. To obtain a simple tight-binding model, we truncate the hopping to the next nearest neighbor which are the minimal interactions needed to explain the physics presented in Fig. 1(c). The momentum space Hamiltonian for a periodic system exhibiting $C_{4v}$ symmetry is thus

$$H = \begin{pmatrix} -4\cos k_x + \cos k_y\left(2 - \cos k_x/\sqrt{2}\right) & 3\sin k_x \sin k_y/\sqrt{2} \\ 3\sin k_x \sin k_y/\sqrt{2} & 2\cos k_x - \cos k_y\left(4 + \cos k_x/\sqrt{2}\right) \end{pmatrix}. \qquad (2)$$

Due to the $C_{4v}$ symmetry, bands are degenerate at $\Gamma$ and $M$. The bands are nondegenerate except for those two high symmetry points (See Appendix A). We note that though we derive the Hamiltonian with the coupled dipole equation, a similar tight-binding Hamiltonian can also describe electronic systems of the same symmetry if p-orbitals ($P_x$ and $P_y$) are dominant in the energy range of interest.

As we are interested in a strip geometry which is periodic along the *x*-direction and finite along the *y*-direction, we provide the projected band structure of the periodic system along the $k_x$ direction as shown in Figs. 2(a-d) with the light gray background (the projected bulk band continuum) as a reference. Here we can focus on the $k_x > 0$ region as the $k_x < 0$ region is simply related by time-reversal symmetry. Now we have a band gap region between the two bands and the band gap closes only at $k_x = 0$ and $k_x = \pi/a$. The bulk polarization $p_y = \int_{BZ} -i\langle u_i | \partial_{k_y} | u_i \rangle dk_y$ for any value of $k_x$ is quantized as $\pi$, and hence a semi-infinite system possesses a topological boundary mode which is

located inside the band gap with dispersion connecting the two band edges[37]. (See Appendix A) The FSE introduces coupling between two such boundary modes located at the $+y$ and $-y$ boundaries if the system is finite.

Then we proceed to investigate the FSE for a finite number of lattice sites ($N_y$) along the $y$-direction. First, we start with an extreme case where $N_y = 1$. This case corresponds to an infinite chain of dipoles. In this case, $P_x$ and $P_y$ dipoles decouple and exhibit positive and negative dispersions, respectively.[38] These two bands cross once at $k_x = \pi/2a$. When $N_y = 2$, i.e., two coupled infinite chains, $P_x$ and $P_y$ dipole couple with each other and there is no pure band with only $P_x$ or $P_y$ component. Instead, one can label the band with either mirror symmetric (even) or antisymmetric (odd). As shown in Fig. 2(b), there are now four bands in total and two of them appear inside the bulk gap region, with one even state (red) and one odd state (blue) for the $P_y$ component. Interestingly, now the red and blue bands cross with each other twice which is also the number of the coupled chains. As we further increase $N_y$ as shown in Figs. 2(c) and 2(d), the gray bands start filling up the projection of the bulk periodic bands, and the red and blue bands approach the dispersion of the boundary modes for a semi-infinite system. Still, the number of nodes is always equal to $N_y$.

To prove that the number of nodes between the even and odd modes is equal to $N_y$, we analytically solve the system. (Proof in Appendix B). We first obtain the eigen energy and eigenstates of the boundary mode for a semi-infinite system as

$$E_e = -\frac{\cos k_x \left[ 6\sqrt{2} \sin k_x + 3 \cos k_x \sin k_x + \xi \left( \cos^2 k_x + 6\sqrt{2} \cos k_x + 16 \right) \right]}{\xi \left( \sqrt{2} \cos k_x + 8 \right) + 3\sqrt{2} \sin k_x} \tag{3}$$

and

$$\mathbf{P}_e = \{p_1, p_2, \cdots, p_i, \cdots\} \begin{pmatrix} 1 \\ i\xi \end{pmatrix}, \tag{4}$$

respectively. In Eq. (4), the subscript labels the number of dipole chain, and the first (second) element inside the parenthesis represent the $P_x$ ($P_y$) component.

$$\xi = \sqrt{(2\sqrt{2} - \cos k_x)/(4\sqrt{2} + \cos k_x)} \tag{5}$$

and

$$p_n = \frac{1}{2^n N\sqrt{4d_1 + d_2^2}} \left[ \left(d_2 + \sqrt{4d_1 + d_2^2}\right)^n - \left(d_2 - \sqrt{4d_1 + d_2^2}\right)^n \right] \tag{6}$$

with $N$ being the normalization constant, $d_1 = -C_1/C_3$, $d_2 = -C_2/C_3$, and

$$\begin{cases} C_1 = \frac{1}{4}\left[\sqrt{2}\cos k_x + 3\sqrt{2}\xi \sin k_x - 4\right] \\ C_2 = 4\cos k_x - E_e \\ C_3 = \frac{1}{4}\left[\sqrt{2}\cos k_x - 3\sqrt{2}\xi \sin k_x - 4\right] \end{cases} \tag{7}$$

To solve for the FSE for the boundary modes in a finite system, we use the eigenfunction of the semi-infinite system as basis. With perturbation theory to the first order, we obtain that the hopping strength between the two boundary modes localized on opposite boundaries is

$$\Delta E = \frac{C_3 \left(p_1 p_{N_y - 1} - p_2 p_{N_y}\right)}{p_1 \left(p_1 + p_{N_y}\right)} \tag{8}$$

Thus, to the first-order approximation, the frequencies of the even and odd boundary modes are $E_e + \Delta E$ and $E_e - \Delta E$, respectively. When $\Delta E = 0$, the hopping strength is zero thus the FSE vanishes, and then there is a nodal point on the even and odd boundary bands. It can be proved that $\Delta E$ is an oscillation function of $k_x$ and exhibits $N_y$ zeros for $k_x \in (0, \pi/a)$. (Detail proof is also provided in Appendix B). To check that the zeros of $\Delta E$ indeed predict the number of nodes, we provide the locations of boundary band nodes (red +) and the zeros of $\Delta E$ (blue square) in Fig. 2(e). The number of the red + exactly equals the number of the blue square for every $N_y$, and the difference comes from the first-order approximation we take. With the increase of $N_y$ where the prediction of $\Delta E$ works better, red + and blue square approach each other.

**III. FILTERING EFFECT**

The boundary modes can be utilized as waveguide channels to guide waves. In our system, the coupling strength between two boundary modes on opposite sides vanishes at these nodes and thus each boundary mode can be confined to one side of the waveguide as it propagates. As the coupling between boundary modes localized on two boundaries is a function of $k_x$ and $N_y$, we can use this unique feature to filter out the component we need. In Fig. 3, we demonstrate this effect. The system consists of four chains ($N_y = 4$) and the wave function is assumed to be initially only on the 4$^{th}$ chain as sketched in Fig. 3(a) with the amplitude given by

$$\psi_{m,4}(t=0) = \exp\left[ik_{x0}ma - \frac{(m-m_0)^2}{w^2}\right]\begin{pmatrix} 1 \\ i\xi(k_{x0}) \end{pmatrix} \quad (9)$$

where we assume a Gaussian package with center wave vector, width and center position given by $k_{x0} = \pi/2a$, $w=10$ and $m_0 = 20$, respectively. $\xi$ is set as $\xi = 0.71$ to match the boundary modes at $k_{x0}$ as given by Eq. (B3). All the lattice sites are set as lossless (orange) except the first column which is absorptive with onsite energy $\gamma_1 = -0.2i$. We solve the time-dependent Schrodinger equation. For simplicity, we set $h=1$ in the simulation. With the propagating of the wave package, the unwanted components are coupled to the first column and get absorbed eventually. Figure 3(b) gives the normalized $k_x$ component as a function of evolution time. At $t=0$, it exhibits a Gaussian shape centered at $k_{x0}$ as given by Eq. (9). With the increase of $t$, the Gaussian shape gradually evolves to two sharp peaks at $k_x a / 2\pi = 0.22$ and 0.28, which are exactly the $k_x$ values at which the two boundary modes decoupled.

**IV. Plasmonic sphere array with this FSE**

Up to this point, we used a tight-binding model with next-nearest neighbor coupling to reveal the physics of such a peculiar FSE. We then demonstrate that such a novel effect exists with full wave simulations when all the coupling terms are considered. As our system only requires the symmetry imposed by the $P_x$ and $P_y$ modes in a square lattice, the FSE discussed herein should be quite universal. Possible candidates are photonic crystals, phononic crystals, cold atoms, and 2D materials with properly chosen orbitals. Below we show that this peculiar type of FSE also exists in a plasmonic sphere array. We consider an array of plasmonic spheres with lattice constant $a = 50\text{nm}$ and the radius of the sphere $r_s = 20\text{nm}$. We employ a Drude model for the sphere: $\varepsilon(\omega) = 1 - \omega_p^2 / \omega(\omega + i\gamma_p)$ with the plasmon frequency $\omega_p = 4eV$ and the damping coefficient $\gamma_p = 0.01eV$. The coupled dipole equation is

$$\mathbf{p}_m = \alpha \left[ \mathbf{E}_m^{\text{ext}} + \sum_{m \neq n} \overset{\leftrightarrow}{\mathbf{W}}(\mathbf{R}_m - \mathbf{R}_n) \mathbf{p}_n \right], \quad (10)$$

where $\mathbf{p}_m$ represents the dipole moment at location $\mathbf{R}_m$, $\mathbf{E}_m^{\text{ext}}$ is the corresponding local external electric field, and $\alpha(\omega) = i3a_1(\omega)/2k_0^3$ is the dynamic dipole polarizability, where $k_0 = \omega/c$ is the wave vector in vacuum with $c$ being the speed of light, $a_1(\omega)$ is the electric dipolar term of the Mie coefficients[39]. $\overset{\leftrightarrow}{\mathbf{W}}$ is the dyadic Green's function,

$$\overset{\leftrightarrow}{\mathbf{W}}(\mathbf{r}) = k_0^3 \left[ A(k_0 r) \mathbf{I}^{3\times 3} + B(k_0 r) \frac{\mathbf{rr}}{r^2} \right] \quad (11)$$

with $\mathbf{I}^{3\times 3}$ being the $3\times 3$ identity matrix and

$$\begin{cases} A(k_0 r) = \left[ (k_0 r)^{-1} + i(k_0 r)^{-2} - (k_0 r)^{-3} \right] e^{ik_0 r} \\ B(k_0 r) = \left[ -(k_0 r)^{-1} - 3i(k_0 r)^{-2} + 3(k_0 r)^{-3} \right] e^{ik_0 r} \end{cases}. \quad (12)$$

Note here, when the wavelength is much larger than the lattice constant, i.e., $k_0 r \ll 1$, $(k_0 r)^{-3}$ terms dominate in $A(k_0 r)$ and $B(k_0 r)$, and $e^{ik_0 r} \approx 1$. Thus we get the quasi-static limit [40-42] and $\overset{\leftrightarrow}{\mathbf{W}}$ reduces to the hopping in Eq. (1).

The coupled dipole equation can be reformulated as $\mathbf{Mp} = \mathbf{E}^{\text{ext}}$, where $\mathbf{M} = \alpha^{-1} - \overset{\leftrightarrow}{\mathbf{W}}$ and the position dependence is omitted for simplicity. One can define the eigen polarizability as $\alpha_{\text{eig}} = \lambda^{-1}$ with $\lambda$ being the eigenvalue of $\mathbf{M}$.[43,44] In a periodic system, $\alpha_{\text{eig}}$ is a function of $\omega$ and the Bloch wave vector. For a passive system, $\text{Im}[\alpha_{\text{eig}}]$ is always positive and exhibits a peak in the presence of a resonance. Moreover, $\text{Im}[\alpha_{\text{eig}}]$ is proportional to the extinction of the driving field, and the width of an extinction peak on $\omega$ is proportional to the mode quality.[43,44] Hence the summation of all the imaginary parts of the eigen-polarizabilities, *i.e.*, $\text{Im}[\sum \alpha_{\text{eig}}]$ can represent the resonance response of the system. In Fig. 4, we set $N_y = 3$ and show $\text{Im}[r_S^{-3} \sum \alpha_{\text{eig}}]$ as a function of $k_x$ and energy. The resonance peak shows the dispersion of such a plasmonic sphere array. In addition to the dispersion, this plot automatically shows the presence of light cone (marked by the black dashed line in Fig. 4) as the interaction of the plasmonic spheres with the free propagating wave is taken into consideration. Except for that, the dispersion is quite similar to Fig. 2(c). Here we can also find three nodes (marked by the red dots) between the middle two bands. Hence we have numerically demonstrated the fact that the nodes of FSE are not limited to the tight-binding Hamiltonian in Eq. (2). The underlying physics should be universal even in the presence of the light cone when we consider $P_x$ and $P_y$ orbitals in a square lattice.

## V. SUMMARY

In summary, we analytically solved a next-nearest-neighbor hopping model to investigate the FSE of boundary modes. The boundary modes of a finite-width strip twist around each other, intersecting at nodes and the number of nodes equals the number of lattices across the strip. This FSE leads to a special type of filtering effect where only the components with wave vectors match those of the nodes preserved. Meanwhile, this peculiar FSE is general and also presents in the plasmonic sphere array. Our model is based on just in-plane dipole orbitals and needs no further assumption, and the FSE discussed here can be found in electronic waves, classical waves and cold atoms. Our work points to the possibility of getting rid of the formation of gaps due to FSE with properly chosen orbitals and lattice, and thus opens a feasible way for integrating multiple waveguides into a compact device.

# Figures:

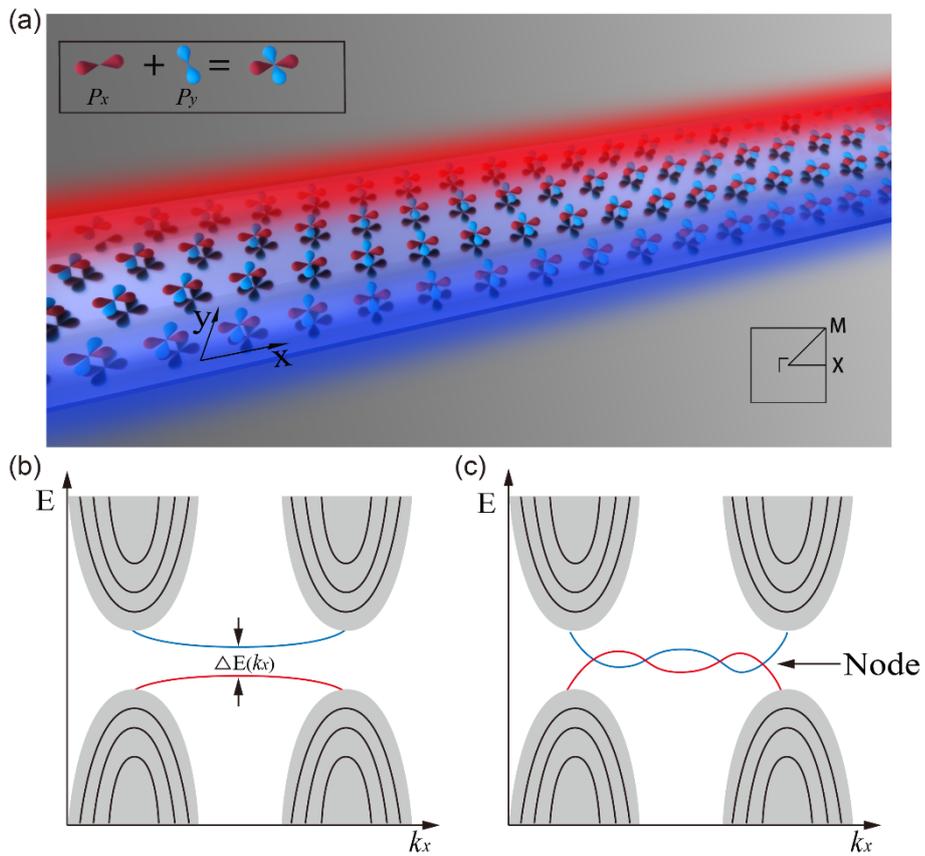

FIG. 1. A peculiar type of FSE. (a) The coupling of two boundary modes (denoted by the red and blue shaded regions) localized on opposite sides of a strip geometry of a square lattice with $P_x$ and $P_y$ orbitals. The lower right inset shows the reciprocal space. (b-c) Sketches of two different types of FSEs, where the FSE is finite for all $k_x$ in (b) and vanishes at several nodes for certain $k_x$s in (c).

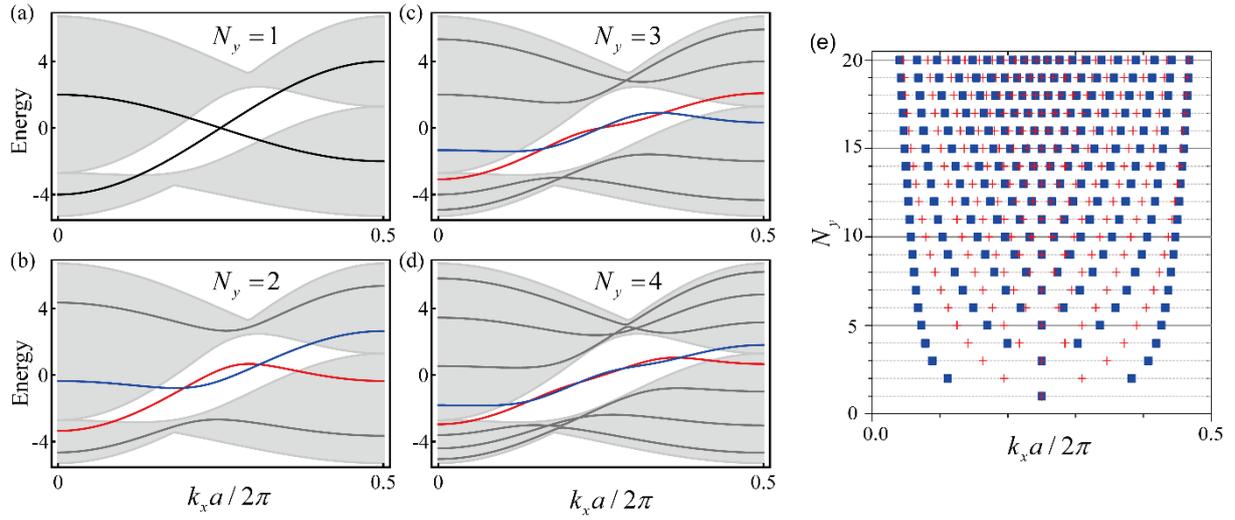

FIG. 2. Band structure of the minimal model. (a-d) The band structures for different $N_y$s, where the gray areas represent the projection of bulk bands, and the red (blue) curves are even (odd) modes with energy predominantly localized at the boundaries. (e) The locations of nodes on the boundary modes (red plus signs) and the zeros of $\Delta E$ [blue square symbols, defined in Eq. (8)] for different $N_y$s.

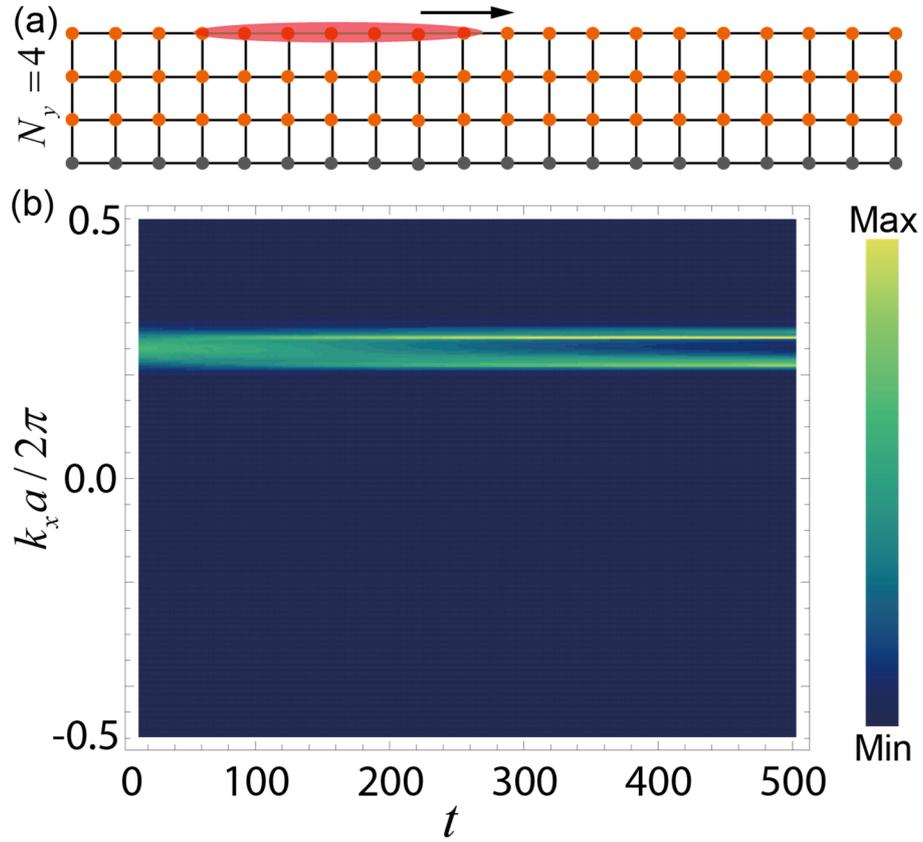

FIG. 3. (a) Sketch of the system in simulation, which is long enough along the *x*-direction and 4 along the *y*-direction. Orange and gray represent lattice sites with no absorption and finite absorption $\gamma_1 = -0.2i$, respectively. (b) The normalized amplitude of wave package in the momentum space as a function of the evolution time *t*.

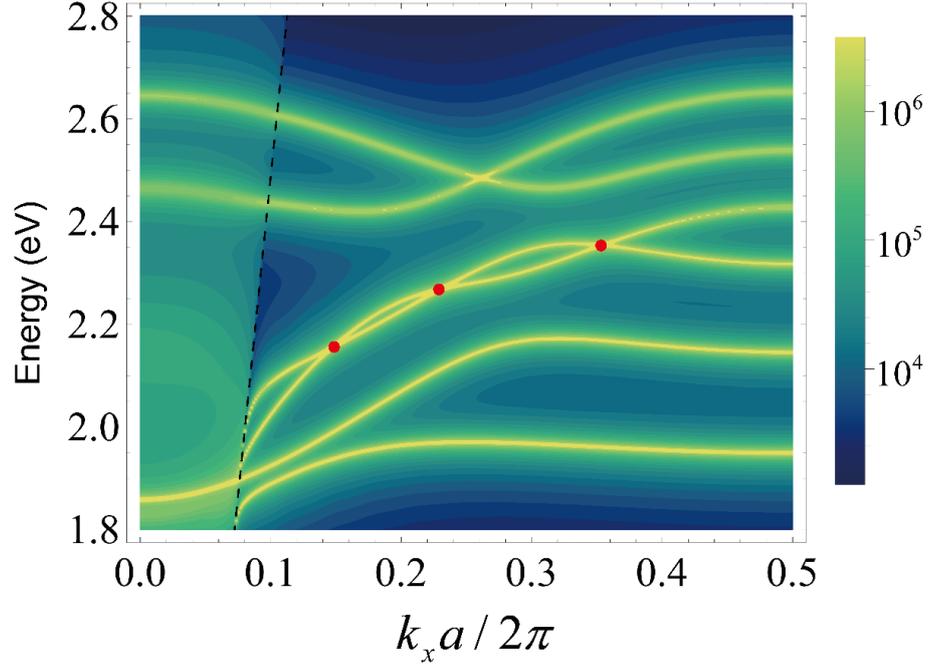

FIG. 4. $\text{Im}\left[r_S^{-3}\sum \alpha_{\text{eig}}\right]$ as a function of $k_x$ and the energy of the electromagnetic waves. Here the black dashed line marks the position of the light cone, and the red dots highlight the positions of nodes on the boundary modes. In this plot, the lattice constant is $a = 50\text{nm}$, the radius of the plasmonic sphere is $r_S = 20\text{nm}$, the plasmonic frequency is $\omega_p = 4eV$ and the damping coefficient is $\gamma_p = 0.01eV$.

**APPENDIX A: BAND STRUCTURE OF THE PERIODIC SYSTEM AND BOUNDARY MODES DISPERSION**

In this section, we provide the band structure of the periodic tight-binding system and the dispersion of the boundary mode for a semi-infinite system. Figure 5(a) gives the band structure of the periodic system. For the dispersion of the boundary mode, we use a finite system with a large enough $N_y$. When $N_y$ is large enough, the coupling between the two boundary modes on opposite sides becomes extremely small, as can be seen in Fig. 5(b) with $N_y = 20$. The red (even mode) and blue curves (odd mode) almost overlap with each other. The dispersion of the boundary mode for a semi-infinite system can then be obtained by taking the average of these two curves[37].

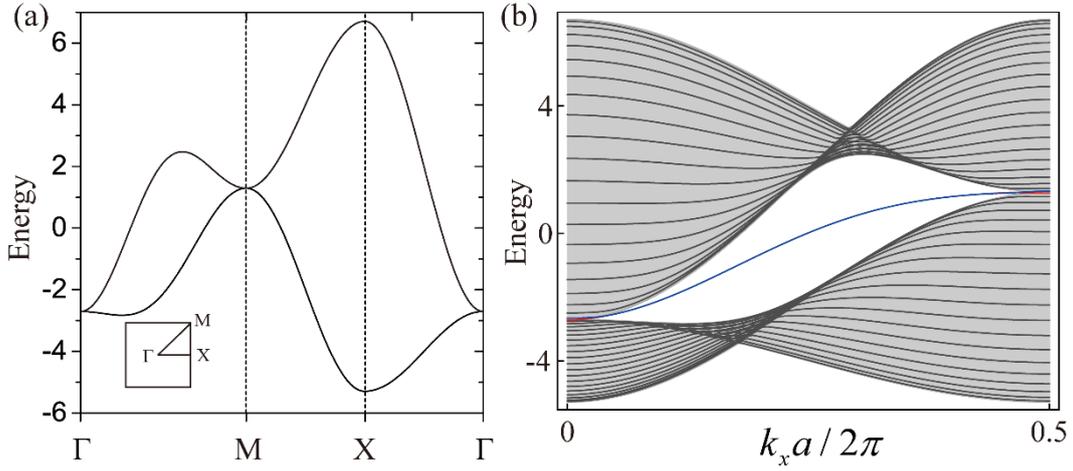

FIG. 5. (a) The band structure of the periodic tight-binding model. (b) Band dispersion with $N_y = 20$, where the gray background represents the projection of the periodic bulk band. The Hamiltonian is provided in Eq. (2) of the main text.

**APPENDIX B: The proof that the number of nodes equals to $N_y$**

In this section, we prove that the number of nodes equals $N_y$ analytically. We first obtain the dispersion and the wave function of the boundary modes for a semi-infinite system. The boundary mode wave function of a semi-infinite system exhibits the following form

$$\mathbf{P}_e = \{p_1, p_2, \cdots, p_i, \cdots\} \begin{pmatrix} 1 \\ i\xi \end{pmatrix}, \tag{B1}$$

where the subscript labels the number of the dipole chain, and the first (second) element inside the parenthesis represents the $P_x$ ($P_y$) component. This boundary mode satisfies the equation

$$H_{\text{semi}}(k_x) \mathbf{P}_e(k_x) = E_e(k_x) \mathbf{P}_e(k_x), \tag{B2}$$

where $H_{\text{semi}}$ is the Hamiltonian of a semi-infinite system and $E_e$ is the eigen energy. After some simple math, we obtain

$$\xi = \sqrt{\left(2\sqrt{2} - \cos k_x\right) / \left(4\sqrt{2} + \cos k_x\right)}, \tag{B3}$$

$$E_e = -\frac{\cos k_x \left[6\sqrt{2} \sin k_x + 3 \cos k_x \sin k_x + \xi\left(\cos^2 k_x + 6\sqrt{2} \cos k_x + 16\right)\right]}{\xi\left(\sqrt{2} \cos k_x + 8\right) + 3\sqrt{2} \sin k_x}. \tag{B4}$$

Meanwhile, the series $\{p_1, p_2, \cdots, p_i, \cdots\}$ satisfies the condition

$$\begin{cases} C_1 p_{i-1} + C_2 p_i + C_3 p_{i+1} = 0 \\ C_2 p_1 + C_3 p_2 = 0 \end{cases}, \tag{B5}$$

with $p_1$ being used for normalization and

$$\begin{cases} C_1 = \frac{1}{4}\left[\sqrt{2} \cos k_x + 3\sqrt{2} \xi \sin k_x - 4\right] \\ C_2 = 4 \cos k_x + E_e \\ C_3 = \frac{1}{4}\left[\sqrt{2} \cos k_x - 3\sqrt{2} \xi \sin k_x - 4\right] \end{cases}. \tag{B6}$$

Equation (B5) provides a recurrence relation for $p_n$, which gives

$$p_n = \frac{1}{2^n N \sqrt{4d_1 + d_2^2}} \left[\left(d_2 + \sqrt{4d_1 + d_2^2}\right)^n - \left(d_2 - \sqrt{4d_1 + d_2^2}\right)^n\right], \tag{B7}$$

where $N$ is the normalization constant, $d_1 = -C_1/C_3$ and $d_2 = -C_2/C_3$. As a preliminary check, we compare Eqs. (B3) and (B4) with numerical results from finite systems as shown in Figs. 6(a) and 6(b), respectively. In Fig. 6(a), the black line is obtained from Eq. (B3), and the blue, magenta and red represent $|p_{1y}/p_{1x}|$ for $N_y = 6, 12, 30$, respectively. It is clear that with the increasing of $N_y$, numerical results approaching that from Eq. (B3). Meanwhile, numerical results deviate from Eq. (B3) near $k_x = 0$ and $\pi/a$ whereat the band gaps close. Figure 6(b) shows the comparison of numerically obtained boundary mode dispersion for a large enough system (blue line) and that from Eq. (B4) (open red disk). They agree with each other perfectly well.

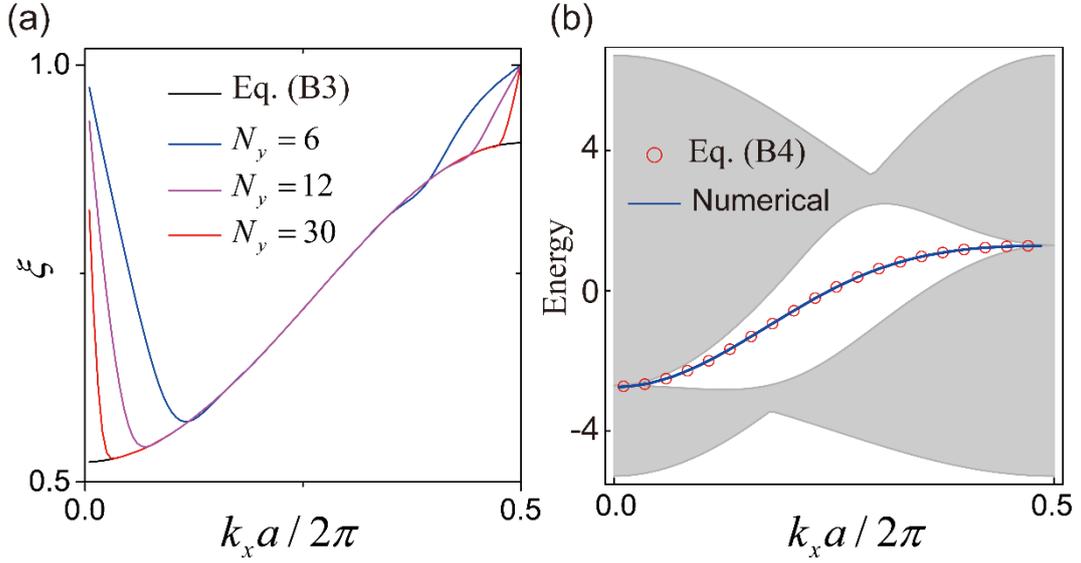

FIG. 6. (a) Comparing of $\xi$ obtained from Eq. (B3) for a semi-infinite system and the numerically obtained value for different $N_y$s. (b) The boundary mode dispersions obtained from Eq. (B4) (open red circle) and the numerical simulations (blue line). Here the gray regions represent the projection of the bulk bands.

The coefficient $2^{-n}$ in Eq. (B7) provides the exponentially decaying feature of $p_n$ with the increasing of $n$. To obtain more detailed features of $p_n$, we need the information of $d_2$ and $\sqrt{4d_1 + d_2^2}$ as functions of $k_x$, which are provided in Fig. 7(a). It can be shown that $d_2$ is a monotonically decreasing function of $k_x$, and $\sqrt{4d_1 + d_2^2}$ is purely imaginary and exhibits a $\sin k_x$-like shape. We can define

$$\theta = \arg\left(d_2 + \sqrt{4d_1 + d_2^2}\right), \tag{B8}$$

and show that $\theta$ is a monotonically increasing function of $k_x$. $\theta = 0$ when $k_x = 0$; $\theta = \pi$ when $k_x a = \pi$. Taking Eq. (B8) into Eq. (B7), we have

$$p_n = \frac{(-2d_1)^{n-1}}{N} \frac{\sin(n\theta)}{\sin\theta}. \tag{B9}$$

We can see that $\{p_1, p_2, \cdots, p_i, \cdots\}$ are oscillating decaying functions of the chain index $i$. Noting that $\theta$ is a monotonically increasing function of $k_x$, $p_n$ should thus also be an oscillation function of $k_x$, which gives the essential reason for the emergence of nodes. And Eq. (B9) proves that $p_n$ possesses $n-1$ zeros for $k_x \in (0, \pi)$. The first few $p_n$ functions are provided in Figs. 7(b) and 7(c). We can clearly see the oscillation feature and observe that $p_n$ exhibits $n-1$ zeros.

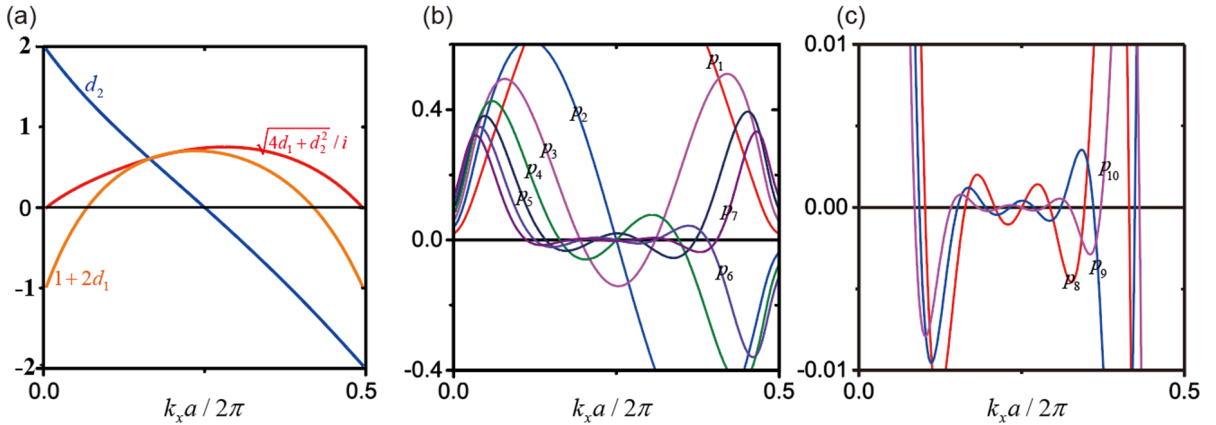

FIG. 7. (a) $\sqrt{4d_1 + d_2^2}/i$ (red), $d_2$ (blue), $1+2d_1$ (orange) as a function of $k_x$. (b), (c) $p_n$ as functions of $k_x$, where different colors for different $n$. $p_n$ is an oscillation function of $k_x$ and has $n-1$ zeros.

For a finite system, the wave function of the boundary mode should be

$$\mathbf{P}_{\text{finite}} = \left( \{p_1, p_2, \cdots, p_i, \cdots\} \pm \{p_{N_y}, p_{N_y-1}, \cdots, p_{N_y+1-i}, \cdots\} \right) \begin{pmatrix} 1 \\ i\xi \end{pmatrix}. \tag{B10}$$

The wave function satisfies

$$\begin{pmatrix} \tilde{C}_2 & C_3 & & & \\ C_1 & \tilde{C}_2 & C_3 & & \\ & \ddots & \ddots & \ddots & \\ & & C_1 & \tilde{C}_2 & C_3 \\ & & & C_1 & \tilde{C}_2 \end{pmatrix} \begin{pmatrix} p_1 \pm p_{N_y} \\ p_2 \pm p_{N_y-1} \\ \vdots \\ p_{N_y-1} \pm p_2 \\ p_{N_y} \pm p_1 \end{pmatrix} = 0, \tag{B11}$$

where $\tilde{C}_2 = 4\cos k_x + E_e \pm \Delta E$ with $\Delta E$ being half of the energy difference between the even and odd modes. $\Delta E$ also equals the hopping strength between the boundary modes localized on different boundaries. If we assume that $\Delta E$ is a small number and can thus treat it as a perturbation. Then to the first order of $\Delta E$, we have

$$\Delta E = \frac{C_3 \left( p_1 p_{N_y-1} - p_2 p_{N_y} \right)}{p_1 \left( p_1 + p_{N_y} \right)}. \tag{B12}$$

Considering the fact that $C_3 / p_1 \left( p_1 + p_{N_y} \right)$ exhibit no zeros, and combined with Eq. (B7), we find that the zeros of $\Delta E$ is the same as the function (the numerator without some trivial constants)

$$\Delta \tilde{E}_n = (1 + 2d_1) d_2 \left[ \left( d_2 + \sqrt{4d_1 + d_2^2} \right)^n - \left( d_2 - \sqrt{4d_1 + d_2^2} \right)^n \right]$$
$$+ \sqrt{4d_1 + d_2^2} \left[ \left( d_2 + \sqrt{4d_1 + d_2^2} \right)^n + \left( d_2 - \sqrt{4d_1 + d_2^2} \right)^n \right], \tag{B13}$$
$$= i(-4d_1)^{n+1} \left[ -(1 + 2d_1) \cos\theta \sin(n\theta) + \sin\theta \cos(n\theta) \right]$$

where Eq. (B8) is used in the last step. Noting that $d_1 < 0$ [see Fig. 7(a)], the zeros of $\Delta \tilde{E}_n$ are the solutions of the following equations,

$$1 + 2d_1 = F_n \equiv \tan\theta \cot(n\theta) \tag{B14a}$$

or

$$\sin(n\theta) = \sin\theta = 0 \tag{B14b}$$

Equation (B14b) is satisfied when $n$ is an odd number and $k_x a = \pi/2$. For Eq. (B14a), we first notice that $F_n$ exhibits the following features as shown in Fig. 8:

a. $F_n$ is a monotonic decreasing function of $\theta$ for $\theta < \pi/2$ and a monotonic increasing function of $\theta$ for $\theta > \pi/2$.

b. $F_n$ exhibits $n-1$ singular points at $\theta = m\pi/n$, where $m \in \{1, 2, \cdots, n-1\}$. Meanwhile $F_n$ diverges when $\theta$ approaching these singular points.

Taking the value of $1+2d_1$ as given in Fig. 7(a) into consideration, we can conclude that there are in total $n$ zeros for $\Delta\tilde{E}_n$ as a function of $k_x$ for $k_x a \in (0, \pi)$.

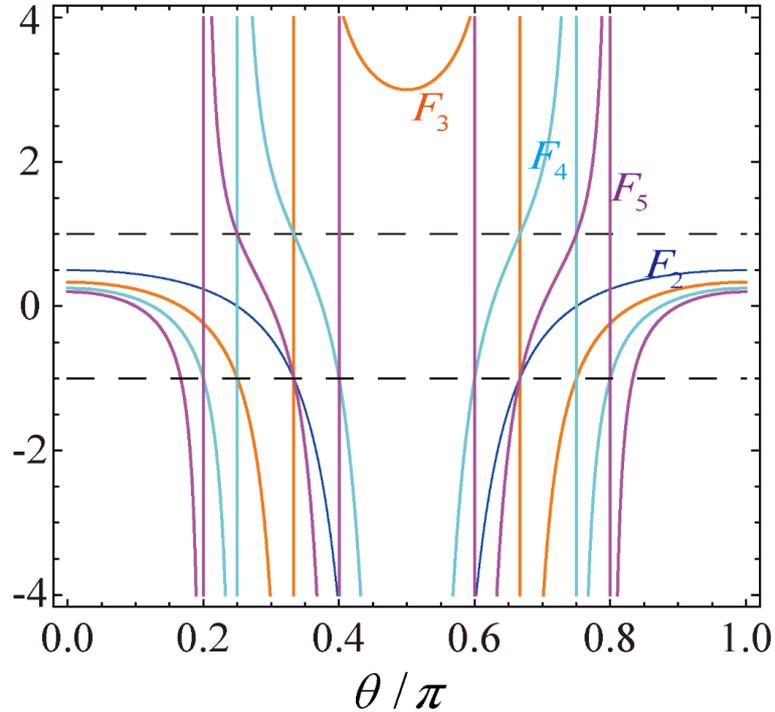

FIG. 8. $F_i$ defined in Eq. (B14a) with $i \in \{2, 3, 4, 5\}$ as a function of $\theta$. The black dashed lines are $y=1$ and $y=-1$ for reference.

# References:

[1] R. Ashoori, Nature **379**, 413 (1996).

[2] U. Banin, Y. Cao, D. Katz, and O. Millo, Nature **400**, 542 (1999).

[3] C. Kittel, *Introduction to Solid State Physics* (Wiley, 2004).

[4] J. D. Joannopoulos, S. G. Johnson, J. N. Winn, and R. D. Meade, *Photonic Crystals: Molding the Flow of Light (Second Edition)* (Princeton University Press, 2008).

[5] J. L. O'brien, A. Furusawa, and J. Vučković, Nat. Photonics **3**, 687 (2009).

[6] M. Saffman, T. G. Walker, and K. Mølmer, Rev. Mod. Phys. **82**, 2313 (2010).

[7] A. Imamog̅lu, D. D. Awschalom, G. Burkard, D. P. DiVincenzo, D. Loss, M. Sherwin, and A. Small, Phys. Rev. Lett. **83**, 4204 (1999).

[8] J. I. Cirac, P. Zoller, H. J. Kimble, and H. Mabuchi, Phys. Rev. Lett. **78**, 3221 (1997).

[9] T. Pellizzari, S. A. Gardiner, J. I. Cirac, and P. Zoller, Phys. Rev. Lett. **75**, 3788 (1995).

[10] R. Horodecki, P. Horodecki, M. Horodecki, and K. Horodecki, Rev. Mod. Phys. **81**, 865 (2009).

[11] N. Akopian, N. H. Lindner, E. Poem, Y. Berlatzky, J. Avron, D. Gershoni, B. D. Gerardot, and P. M. Petroff, Phys. Rev. Lett. **96**, 130501 (2006).

[12] A. Aspuru-Guzik and P. Walther, Nat. Phys. **8**, 285 (2012).

[13] X.-L. Zhang, F. Yu, Z.-G. Chen, Z.-N. Tian, Q.-D. Chen, H.-B. Sun, and G. Ma, Nat. Photonics **16**, 390 (2022).

[14] J. M. Arrazola *et al.*, Nature **591**, 54 (2021).

[15] H. A. Haus and K. F. Šipilov, *Waves and Fields in Optoelectronics* (Prentice-Hall, 1984).

[16] T. Song, H. Chu, J. Luo, Z. Cao, M. Xiao, R. Peng, M. Wang, and Y. Lai, Phys. Rev. X **12**, 011053 (2022).

[17] M. Z. Hasan and C. L. Kane, Rev. Mod. Phys. **82**, 3045 (2010).

[18] X.-L. Qi and S.-C. Zhang, Rev. Mod. Phys. **83**, 1057 (2011).

[19] T. Ozawa *et al.*, Rev. Mod. Phys. **91**, 015006 (2019).

[20] G. Ma, M. Xiao, and C. T. Chan, Nat. Rev. Phys. **1**, 281 (2019).

[21] F. D. M. Haldane and S. Raghu, Phys. Rev. Lett. **100**, 013904 (2008).

[22] Z. Wang, Y. D. Chong, J. D. Joannopoulos, and M. Soljačić, Phys. Rev. Lett. **100**, 013905

(2008).

[23] Z. Wang, Y. Chong, J. D. Joannopoulos, and M. Soljačić, Nature **461**, 772 (2009).

[24] Z. Yu, G. Veronis, Z. Wang, and S. Fan, Phys. Rev. Lett. **100**, 023902 (2008).

[25] X. Ni, M. Li, M. Weiner, A. Alù, and A. B. Khanikaev, Nat. Commun. **11**, 1 (2020).

[26] B. Zhou, H.-Z. Lu, R.-L. Chu, S.-Q. Shen, and Q. Niu, Phys. Rev. Lett. **101**, 246807 (2008).

[27] M. Ezawa and N. Nagaosa, Phys. Rev. B **88**, 121401(R) (2013).

[28] J. Linder, T. Yokoyama, and A. Sudbø, Phys. Rev. B **80**, 205401 (2009).

[29] H.-Z. Lu, W.-Y. Shan, W. Yao, Q. Niu, and S.-Q. Shen, Phys. Rev. B **81**, 115407 (2010).

[30] V. Krueckl and K. Richter, Phys. Rev. Lett. **107**, 086803 (2011).

[31] L.B. Zhang, F. Cheng, F. Zhai, and K. Chang, Phys. Rev. B **83**, 081402(R) (2011).

[32] W. Shockley, Physical review **56**, 317 (1939).

[33] I. Tamm, Phys. Z. Sowjetunion. **1**, 733 (1932).

[34] C.-X. Liu, H.J. Zhang, B. Yan, X.-L. Qi, T. Frauenheim, X. Dai, Z. Fang, and S.-C. Zhang, Phys. Rev. B **81**, 041307(R) (2010).

[35] V. Markel, J. Mod. Opt. **40**, 2281 (1993).

[36] J. D. Jackson, *Classical Electrodynamics Third Edition* (Wiley, 1998).

[37] X. Huang, M. Xiao, Z.-Q. Zhang, and C. T. Chan, Phys. Rev. B **90**, 075423 (2014).

[38] M. L. Brongersma, J. W. Hartman, and H. A. Atwater, Phys. Rev. B **62**, R16356 (2000).

[39] C. F. Bohren and D. R. Huffman, *Absorption and Scattering of Light by Small Particles (Wiley science paperback series)* (Wiley-VCH, 1998).

[40] M. L. Brongersma, J. W. Hartman, and H. A. Atwater, Phys. Rev. B **62**, R16356 (2000).

[41] D. Han, Y. Lai, J. Zi, Z.-Q. Zhang, and C. T. Chan, Phys. Rev. Lett. **102**, 123904 (2009).

[42] L. Wang, R.-Y. Zhang, M. Xiao, D. Han, C. T. Chan, and W. Wen, New J. Phys. **18**, 103029 (2016).

[43] K. H. Fung and C. T. Chan, Opt. Lett. **32**, 973 (2007).

[44] Y.-R. Zhen, K. H. Fung, and C. T. Chan, Phys. Rev. B **78**, 035419 (2008).